In memory of Neil Ashcroft (1938–2021)

# High-temperature superconductivity in hydrides

I. A. Troyan[1], D. V. Semenok[2], A. G. Ivanova[1], A. G. Kvashnin[2], D. Zhou[2], A. V. Sadakov[3], O. A. Sobolevskiy[3], V. M. Pudalov[3,4], I. S. Lyubutin[1], A. R. Oganov[2]

[1] Shubnikov Institute of Crystallography, Federal Scientific Research Center Crystallography and Photonics, 59 Leninsky Prospekt, Moscow 119333

[2] Skolkovo Institute of Science and Technology, Skolkovo Innovation Center, 30 Bolshoy Boulevard, bld. 1, Moscow 121205

[3] P. N. Lebedev Physical Institute, Russian Academy of Sciences, 53 Leninsky Prospekt, Moscow 119991

[4] National Research University Higher School of Economics, 11 Pokrovsky Bulvar, Moscow 109028

## Abstract

Over the past six years (2015–2021), many superconducting hydrides with critical temperatures $T_C$ up to 250 K, which are currently record highs, have been discovered. Now we can already say that a special field of superconductivity has developed. This is hydride superconductivity at ultrahigh pressures. For the most part, the properties of superhydrides are well described by the Migdal–Eliashberg theory of strong electron–phonon interaction, especially when anharmonicity of phonons is taken into account. We investigate the isotope effect, the effect of the magnetic field (up to 60-70 T) on the critical temperature and critical current in the hydride samples, and the dependence of Tc on the pressure and the degree of doping. The divergences between the theory and experiment are of interest, especially in the regions of phase stability and in the behavior of the upper critical magnetic fields at low temperatures. We present a retrospective analysis of data of 2015-2021 and describe promising directions for future research of hydride superconductivity.

## Contents

# 1. Introduction

Superconductivity (SC) is the property of some materials to possess strictly zero electrical resistance at temperatures below a certain critical temperature $T_C$. More than 100 years of research on this phenomenon have not yet fully revealed the engineering and technical potential of the applications of superconductivity [1,2]; and the microscopic mechanisms of superconductivity are still being discussed. Despite the enormous effect that can be expected from the use of superconductors in various fields of engineering and technology, the scope of their actual practical application is still limited due to high costs for cooling significantly below the critical temperature, as well as due to technical difficulties and high cost of manufacturing multilayer materials of multicomponent composition. Since the discovery of $HgBa_2CaCuO_{6+x}$ in 1993 ($T_C$ = 133 K) [3,4], the search for new, more efficient higher-temperature superconductors at ambient pressure has not yet yielded new results. The microscopic mechanism of superconductivity in cuprates remains an unsolved problem in the theory that impedes the search for new superconductors with higher $T_C$.

There is still no microscopic theory of superconductivity of cuprates, which makes it difficult to predict the search for new superconductors with higher $T_C$. Nevertheless, under high pressure conditions, new superconductors with record $T_C$ were predicted and then experimentally obtained. These are binary polyhydrides with anomalously high hydrogen content such as $Im\bar{3}m$-$H_3S$ ($T_C$ = 203 K) [5,6] and $LaH_{10}$ ($T_C$ = 250-260 K) [7,8]. These results are important not only due to the achieved record $T_C$ values. They clearly demonstrate the prospect of obtaining superconductivity close to room temperature, and also indicate the potential for the mechanism of superconducting pairing through electron-phonon interaction, which was previously underestimated. The search for even higher-temperature superconducting compounds requires a transition to ternary and more complex hydrides, which dramatically increases the variety of possible compounds that are virtually impossible to enumerate by a blind experimental search.

Recent (2005–2015) advances in computational materials science, as well as the prediction of the formation of chemical compounds under extreme pressures of tens and hundreds of gigapascals (GPa) have changed approaches to finding new superconductors. Evolutionary algorithms have reached a high level of predictive accuracy in determining new crystal structures of inorganic compounds and are less costly than "blind" experimental sampling. One of the best methods for predicting thermodynamically stable compounds is USPEX algorithm [9-12].

It has been used to achieve important results in obtaining new superhard materials [13], the first high-temperature superconducting hydrides ($H_3S$ [15, 5, 14], $Si_2H_6$ [16]), and magnetic and electronic materials. In the case of polyhydrides synthesized at high pressures, the *ab initio* methods allow one to establish the structure of the hydrogen sublattice, which cannot be done by X-ray diffraction. The results of a structural search can be verified by measuring the critical temperature of superconductivity in hydrides, since large $T_C$ > 100–200 K are usually associated with a highly symmetrical hydrogen sublattice to achieve the desired parameters of the electron–phonon interaction.

After the discovery of superconductivity in sulfur hydride $H_3S$ in 2015 [5], the next research milestone was the experimental work by Z. Geballe et al. (2018) [17], in which the authors succeeded in synthesizing the previously predicted $LaH_{10}$ superhydride [18,19] at a pressure of 175 GPa. About a year later, superconductivity at 250 K was revealed in the newly found $LaH_{10}$ [7]. Thus, in terms of critical temperature and critical magnetic field lanthanum decahydride has surpassed all cuprate-based compounds found in the last 33 years (since 1986). Only two years passed between the prediction and the discovery of the new record superconductor. Such a minimal time period illustrates the progress (Fig. 1a) achieved in computational materials science and experimental ultra-high pressure techniques using diamond anvils cells. However, lanthanum hydrides have not yet been studied well enough. Higher hydrides LaH$_{10}$ (*C*2/*m*, *R*$\bar{3}$*m*, *Fm*$\bar{3}$*m*, *P*6$_3$/*mmc*) and *P*4/*nmm*-LaH$_{11}$, as well as *Pm*$\bar{3}$*m*-LaH$_{12}$ (at 167 GPa) have been obtained experimentally; multiple superconducting transitions have been observed experimentally in lower lanthanum polyhydrides LaH$_x$ (*x* < 10) [20,21].

Over the past six years (2015–2021, Fig. 1b), a large number of various superhydrides have been synthesized. Among them there are both non-superconducting (for example, FeH$_5$ [22,23], magnetic neodymium superhydrides NdH$_7$ and NdH$_9$ [24], cubic and hexagonal praseodymium hydrides PrH$_9$ [25]) and superconducting (uranium hydrides UH$_7$, UH$_8$, and UH$_9$ stable at record-low pressures [26], thorium polyhydrides ThH$_9$ and ThH$_{10}$ (*T*$_C$ = 161 K [27]), cerium hydrides CeH$_9$ and CeH$_{10}$ (*T*$_C$ ~ 110 K [28,29]), and yttrium hydrides YH$_6$ and YH$_9$ (*T*$_C$ = 224 [30] and 243 K [31])). Most of these compounds were first predicted theoretically and then obtained experimentally, which proves the efficiency of the computational search for thermodynamically stable compounds based on evolutionary algorithms and density functional theory.

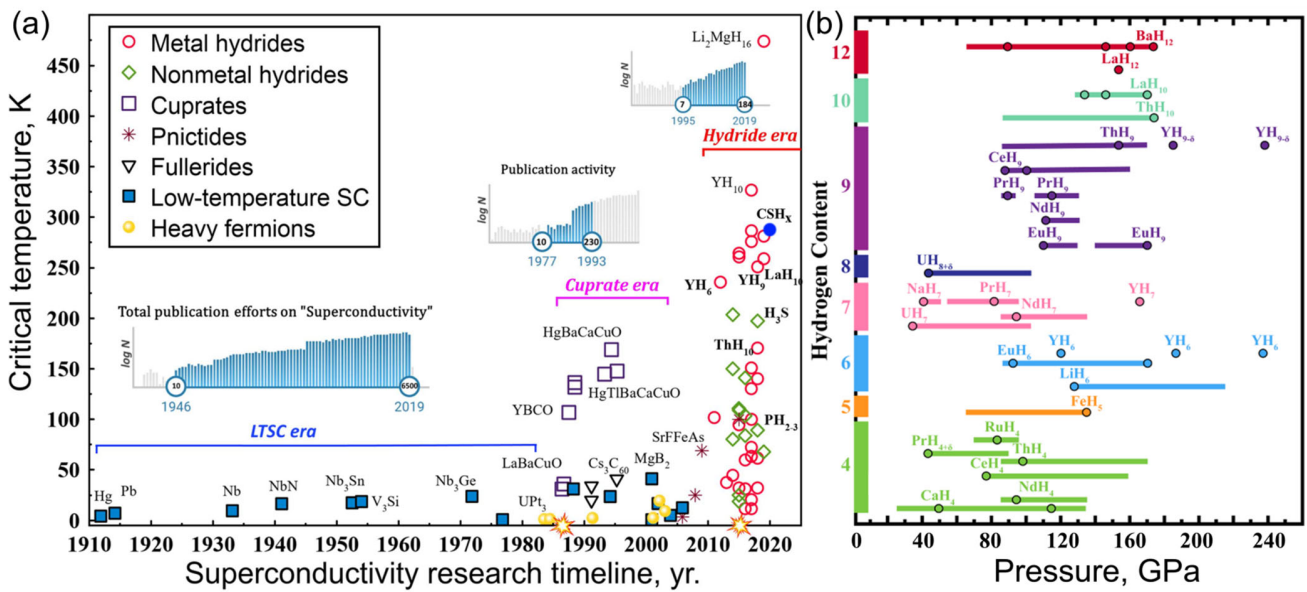

**Figure 1.** (color online). (a) Time scale of superconductors discovery and their critical temperature of superconducting transitions. Metal hydrides are shown by red circles, green diamonds indicate covalent hydrides of nonmetals, asterisks show iron-containing pnictides, hollow squares are cuprate compounds, filled squares are low-temperature superconductors described mainly by the Bardin–Cooper–Shrieffer–Migdal–Eliashberg theory, yellow shaded circles show heavy-fermion systems, and triangles are fullerides. The insets show the number of publications on the corresponding topics on a logarithmic scale (according to Semantic Scholar). (b) Stability regions (in GPa) and hydrogen content per 1 atom of metal (up to *x* = 12) in the best known superhydrides investigated over 2015–2021.

Since the possibilities of formation of high-temperature superconductors among binary hydrides have already been sufficiently studied, the focus is now on ternary systems. The calculations using artificial intelligence algorithms, in particular neural networks, point to ternary hydrides as more promising both in terms of critical temperature and in terms of reduction of the synthesis pressure [32-36]. Thus, in 2020, it was discovered that at a pressure of several gigapascals, a mixture of methane and sulfur hydride H$_2$S photochemically forms a molecular compound. Under further compression, it transforms to a unique material, which is either carbon-doped H$_3$S or an organic compound CSH$_x$. This material demonstrates a sharp drop in resistance at 270 GPa, which, according to Ref. [37], corresponds to a superconducting transition at +15 ºC.

Computer simulation of metal superhydrides shows that at pressures up to 200–300 GPa (at higher pressures, experimental studies of superconductivity are still difficult), the maximum critical temperatures of superconductivity are achieved in hydrides of groups 2 and 3 elements, such as Ca, Sr, Sc, Y, La, Ac, Hf, Zr, Th,

Ce, and Mg, containing 6–10 hydrogen atoms per metal atom. At the moment, the experimental search for promising hydride superconductors is mainly limited to this set of elements and their combinations [36]. For example, calculations show that in the Li–Mg–H system, a clathrate hydride Li$_2$MgH$_{16}$ with $T_C$ above 400 K may exist [32]. However, it is metastable and unlikely to be obtained in experiment.

For the Ca–Y–H and Ca–Mg–H systems, cubic hexahydrides $Pm\bar{3}m$-CaYH$_{12}$ and CaMgH$_{12}$ with critical temperatures of 240–260 K were predicted. Ternary hydrides of lanthanum–cerium, lanthanum–thorium, lanthanum–yttrium, lanthanum–boron [38], and potassium–boron [39] are expected to demonstrate exceptionally high stability (stabilization pressure from 12 to 60 GPa) and critical temperatures of superconductivity above 100 K, bringing us closer to a discovery of a new class of hydrogen-bearing compounds which would be stable under normal conditions. Thus, a great deal of work is to be done to synthesize ternary superhydrides and investigate their properties. In this article, we try to outline future research in this field.

## 2. Classes of polyhydrides

In hydrides, hydrogen can be in different forms: molecular (e.g., LiH$_6$), ionic (KH), and atomic (YH$_6$). According the type of bond "element–hydrogen" hydrides can be divided into covalent (H$_3$S, SnH$_4$), ionic (AlH$_3$), metallic (LaH$_{10}$), and mixed (molecular metal BaH$_{12}$) [40-42]. Also, a subclass of magnetic compounds can be set apart among metal hydrides. For example, magnetic ordering is expected in the hydrides of neodymium NdH$_9$ [24], europium EuH$_9$ [43], samarium SmH$_9$, and of many other lanthanoids. The simultaneous realization of superconductivity in the hydrogen sublattice of hexagonal (e.g., NdH$_9$) or layered (such as FeH$_5$) hydrides and antiferromagnetic ordering in the metal sublattice, can in principle lead to some exotic physical effects typical of cuprates and iron pnictides.

Interest in molecular and mixed superhydrides with a high hydrogen content (pseudo-tetragonal SrH$_{22}$ and BaH$_{21-23}$ [44]) is due to the similarity of their hydrogen sublattices with the structure and properties of some crystal modifications of pure hydrogen (phases II, IV, V). However, the formation of these superhydrides (or hydrogen doped with 4–6% of Sr or Ba) is observed at much lower pressures (100–170 GPa) than those required to obtain the corresponding modifications of pure hydrogen (350–500 GPa). In molecular strontium superhydrides, gradual metallization and a change in optical transparency upon pressure increase from 90 to 160 GPa can be observed [45], whereas barium hydride BaH$_{12}$ [44] demonstrates the emergence of superconductivity and an increase in the critical temperature, in the same way as it was predicted and partially confirmed experimentally for semiconducting and metallic hydrogen [46-49].

The properties of ionic and mixed metal hydrides will probably allow their use as ionic conductors and electrolytes for electrochemical synthesis of hydrides at high pressures, as suggested in ref. [50]. Indeed, calculations show that the hydrogen diffusion rate (~6 × 10$^{-6}$ cm$^2$s$^{-1}$ [51]) at high pressures in hydrides — in Li$_2$MgH$_{16}$ in particular — may be much higher than in known ionic conductors. On the other hand, such a high mobility of hydrogen makes the concept of a specific structure of the hydrogen sublattice in some polyhydrides rather vague. In other words, the hydrogen sublattice can be liquid, while the metal sublattice remains solid.

Covalent hydrides are the most enigmatic class. Because of strong element–hydrogen interactions, the formation of extended polymer chains and various organic groups is possible. The most investigated covalent system is sulfur–carbon–hydrogen. Recent studies of CS$_2$ compression in the diamond anvil cells indicate the formation of complex branched polymers with semiconducting properties [52]. Even five years after the discovery of superconductivity in H$_3$S, more and more hydrides continue to be found in the H – S system [53], which initially seemed quite simple [15]. Moreover, their structure can often be very complex, for example, H$_6$S$_5$ [54]. The observation of a sharp drop in resistance in sulfur–carbon hydride CSH$_x$ [37], interpreted by the authors as superconductivity, has attracted general attention. However, establishing the structure of this compound proved difficult.

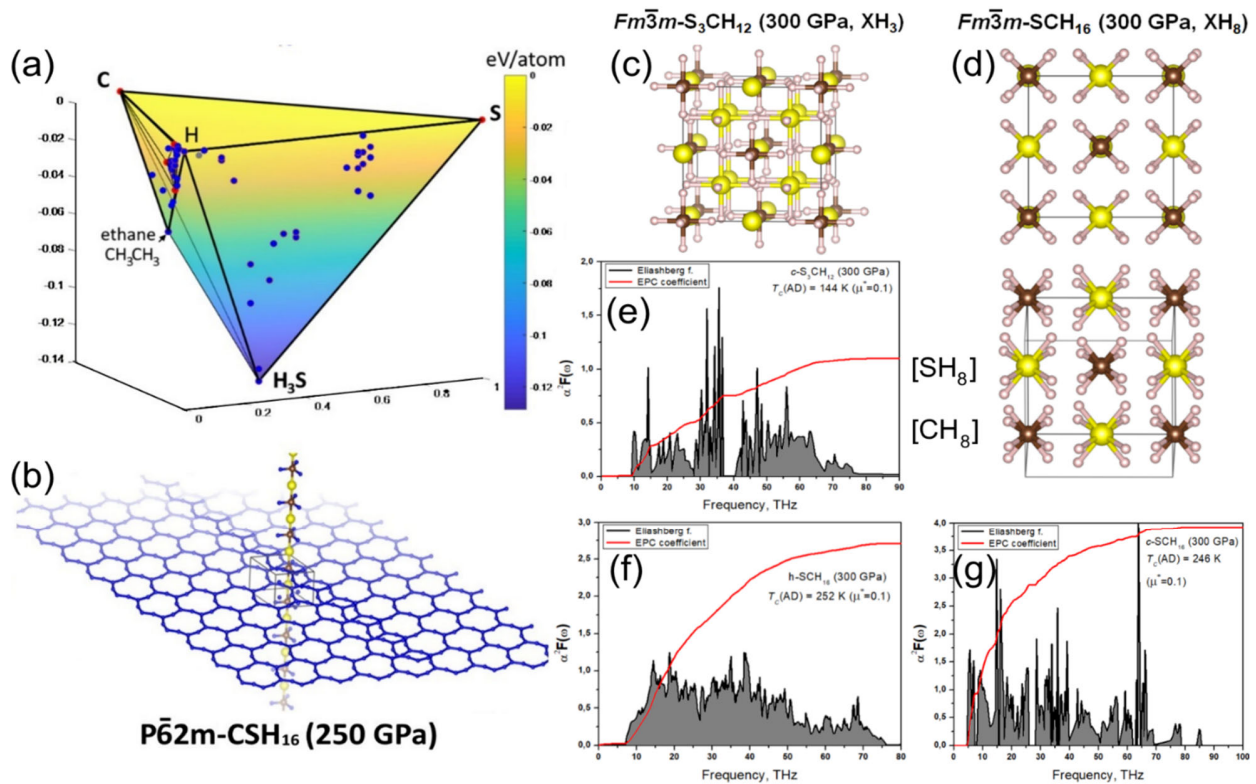

**Figure 2** (color online). (a) Thermodynamic stability diagram of the carbon–sulfur–hydrogen system at 250 GPa, showing the absence of stable ternary compounds; (b – e) structures of unstable C–S–H ternary superhydrides for which high-temperature superconductivity is theoretically possible; (f – h) Eliashberg spectral functions and superconducting transition temperatures $T_C$ calculated by the Allen–Dynes formula (AD [55]). The red curve demonstrates electron-phonon interaction parameter $\lambda(\omega)$.

Extensive theoretical studies of 2020–2021 [56-59] using the standard search for thermodynamically stable phases in the C–S–H system failed to find any convincing candidates up to pressures of 300–350 GPa. Those rather rare phases that could show room temperature superconductivity due to the strong electron–phonon interaction (e.g., $P\bar{6}2m$-CSH$_{16}$ (Fig. 2) or hypothetical $Pn\bar{3}m$-CH$_7$) appear to be highly metastable and should decompose on heating to form previously studied $Im\bar{3}m$-H$_3$S. Almost all organic compounds are metastable with respect to decomposition into simple molecules (CO$_2$, H$_2$O, N$_2$).

However, being in the local minima of the potential energy surface, they are dynamically stable, exist for a long time and are formed by chemical reactions with kinetic control. A similar situation can result from photochemical synthesis at high pressures in the C–S–H system.

In this regard, the actual problem is the refinement of the criteria for selecting structures not by the minimum enthalpy, but by other parameters – such as the best agreement between the chosen structure and the experimental X-ray diffraction pattern, the parameters of the electron-phonon interaction, and various spectra.

To compare with the calculated data, reproducible experimental ones obtained by various methods and in various experiments are required. This is especially important in case of some internal inconsistencies in the experimental data. In particular, this raises doubts that the effect observed in the C-S-H system is superconductivity [58,60,61].

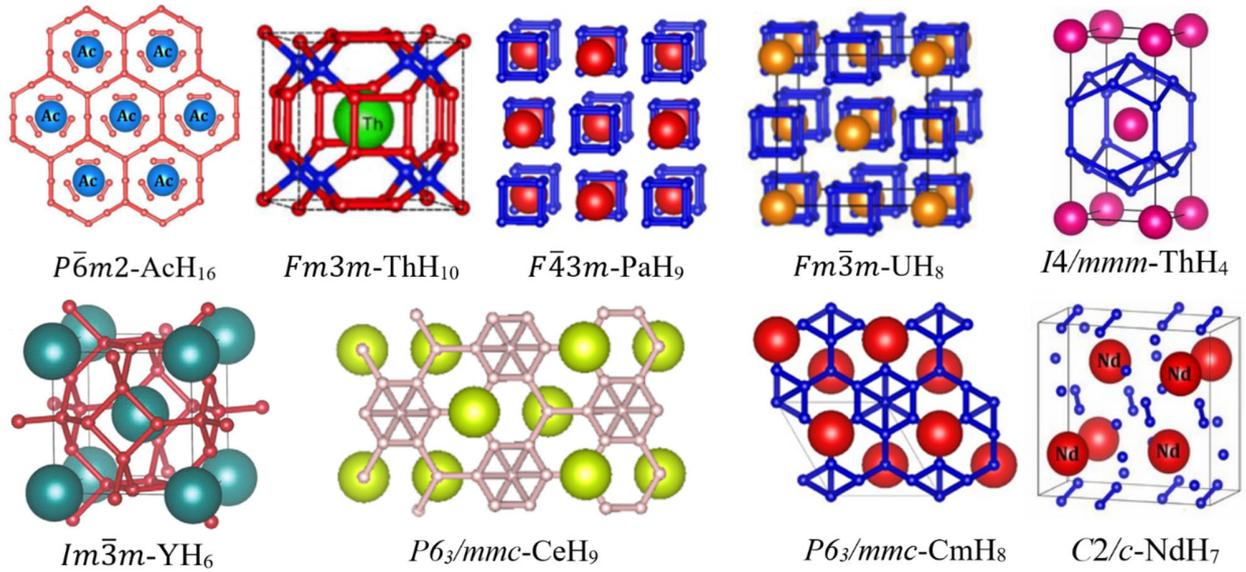

**Figure 3** (color online). Structural motifs of actinoid and lanthanoid superhydrides with a hydrogen atomic sublattice. At high pressures, hexagonal and cubic dense packings of heavy atoms are dominant.

Finally, the simplest and at the same time one of the most important classes are metal superhydrides with atomic hydrogen sublattice (Fig. 3). Such structures are typical metals, whose electrical resistance behavior under ambient conditions is described by the Bloch–Grüneisen formula [62,63]. Below the critical temperature, they exhibit the properties of high-temperature superconductors, have a high density of electronic states at the Fermi level, sometimes with a Van Hove singularity near Fermi level [64-66]. As we have shown previously [36], in this class of compounds there is a certain optimal number of hydrogen atoms per metal atom for achieving the highest critical temperature. The $XH_n$ hydride should have n = 6–10, which formally corresponds to the transfer of ~0.33 electrons per hydrogen atom [19]. Such hydrides are formed at high pressures in reactions of $d^0$–$d^2$ metals with hydrogen and usually have cubic and hexagonal close-packed structures. We have recently found [36] that in terms of a thermodynamics, it is often more favorable for these structures to be slightly distorted. However, this distortion is difficult to detect using existing experimental methods. As pressure decreases, at first more and more deviations from the ideal close-packed packed structure are observed, the critical temperature decreases smoothly. Then the loss of some hydrogen, changes in the composition and symmetry of the structure, and a sharp drop in $T_C$ occur. Thus, the $T_C(P)$ diagram usually [5,7] has the form of a bell or an asymmetrical parabola: $T_C$ decreases in both directions from the maximum either as pressure increases (the electron–phonon interaction constant λ decreases due to "quenching" of the phonon modes) or as it decreases (causing lattice distortion and compound decomposition).

Highly symmetric superhydrides formed by *f*-elements (Pr, Nd, Sm, U, Pu, Am, etc.) and highly symmetric superhydrides (e.g., $PrH_9$ [25], $EuH_9$ [43], $NdH_9$ [24], $UH_7$ [26,67]), do not possess pronounced superconducting properties due to the Cooper pair scattering with spin reversal at paramagnetic centers [68]. Moreover, small additions of *f*-elements effectively suppress superconductivity in hydrides of $d^0$–$d^2$ elements ($LaH_{10}$, $YH_6$), almost without changing their structure [45], which can be used to study the magnetic phase diagram of superhydrides down to the lowest temperatures. To understand the mechanism of superconductivity in hydrides, it is important that small impurities of nonmagnetic elements (such as C, B, N, Al) practically do not affect the critical temperature of hydrides, whereas the introduction of paramagnetic centers (e.g., Nd) dramatically reduces $T_C$.

# 3. Methods for studying metal polyhydrides

X-ray diffraction of materials in high-pressure diamond anvil cells (DAC) using synchrotron radiation is still the main method of determining the crystal structure of hydrides [69]. Synchrotron beam focusing technology has reached submicron resolution. This allows the study of samples that are a few microns or even a few hundred nanometers in size, sandwiched between diamond anvils [70]. Single crystal X-ray diffraction at megabar pressures is becoming increasingly popular [71,72]. This method requires specially shaped diamond anvils with a wide aperture (70–80º). It is used when microcrystals of a certain size (0.25–2 μm) can be grown during a series of laser "annealing" cycles of hydride samples in DAC. Although intense X-ray radiation is dangerous for diamond cells at pressures above 200 GPa because of the risk of anvil cracking [73], no other instrumental methods can provide information of comparable importance.

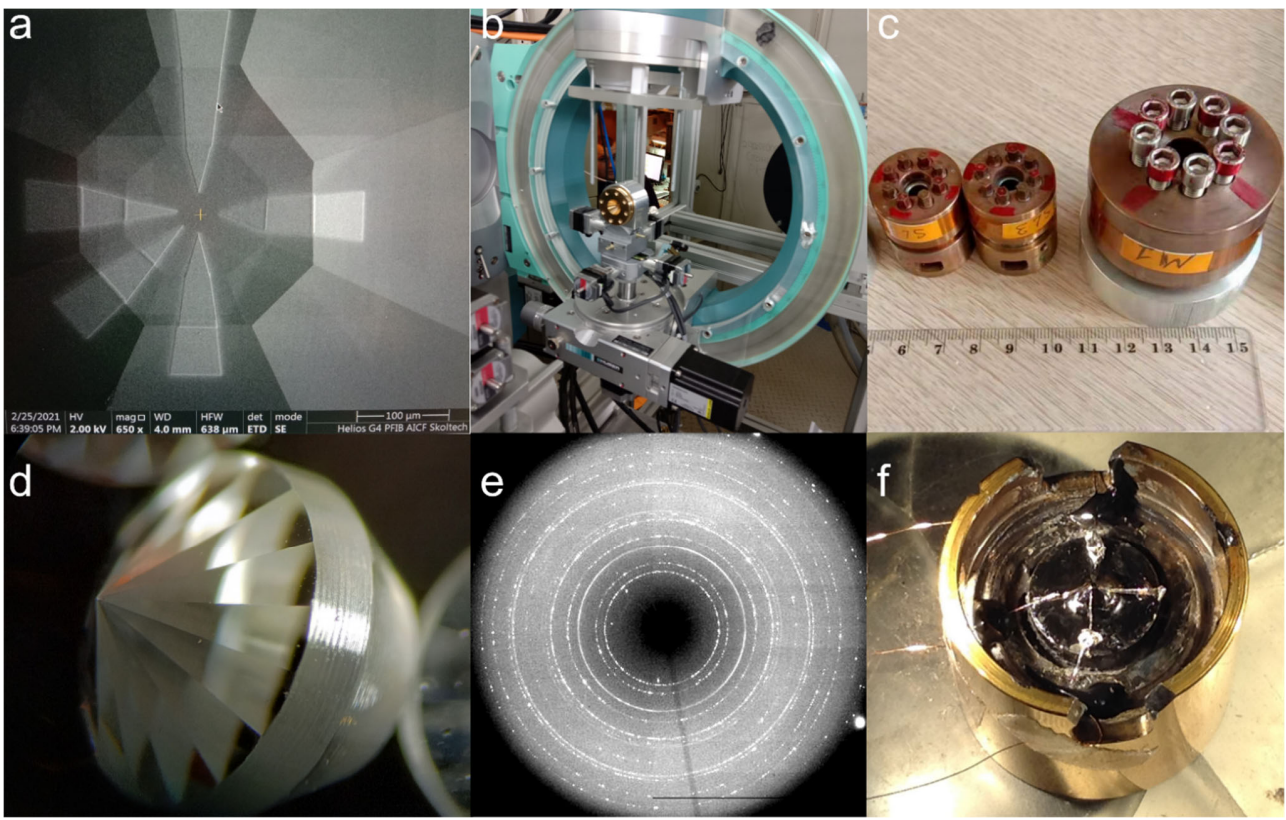

**Figure 4** (color online). Elements of high-pressure diamond anvil cells and experimental technique. (a) System with five platinum electrodes deposited by a focused ion beam to the surface of a diamond anvil. (b) High-pressure cell mounted on a goniometer for synchrotron imaging (Kurchatov synchrotron radiation source). (c) High-pressure bronze cells of various types and sizes. (d) Diamond anvil with a conical base made in Technological Institute of Superhard and New Carbon Materials (TISNCM). (e) X-ray diffraction pattern of thorium superhydride $ThH_{10}$ obtained at ID27 beamline of the European Synchrotron Radiation Facility in Grenoble. (f) Open high-pressure cell with a four-electrode system and copper wires to study the electrical transport characteristics of hydrides.

X-ray diffraction (Figure 4 b,e) allows one to determine the hydrogen content indirectly, using the equation of state of substance investigated in a certain pressure range. Based on the results obtained, one can make assumption on the structure of the hydrogen sublattice. To do this, it must be compared with the most thermodynamically favorable structures found by evolutionary algorithms coupled with density functional theory (DFT). While X-ray diffraction patterns are easily calculated for a given crystal structure, the calculations of the critical temperature of superconductivity, Raman spectra, or sample reflectance are time consuming and highly sensitive to technical parameters, including atomic pseudopotentials. Those are more suitable for confirming the presumed structure but not for finding it "from scratch".

The main methods for characterizing hydrides at pressures above 250–300 GPa are Raman and infrared spectroscopy of semiconductor phases, optical reflection, and measurements of the electrical transport properties in magnetic fields. However, these methods successfully developed for studying metallization of hydrogen [46-49], provide limited information and require time-consuming calculations. Recently, attempts have been made to compare the Raman signals detected in some cases from samples in DAC with the expected spectra of metallic superhydrides [74]. This is rather risky, because calculations of resonance Raman spectra for metals are complex. Whereas dielectric micro-impurities and nanofilms of oxides, hydroxides, organic resins, and other compounds used in the design of DAC give comparable to Raman or stronger response than the expected signal from metal hydrides. To date, there are no published data on systematic studies of the correspondence between Raman spectra and X-ray diffraction results for metal hydrides at high pressures.

Reflection/transmission spectroscopy in the infrared and visible regions enable one to determine the bandgap of compounds, the magnitude and temperature dependence of the superconducting gap, and to compare the calculated electronic band structure with experimental data [48,75,76]. An important requirement for the realization of these methods is the purity of diamonds, low content of nitrogen and other impurities, low concentration of defects, and low luminescence. All these requirements are fulfilled for synthetic diamonds produced by the HPHT (High Temperature – High Pressure) and CVD (Chemical Vapor Deposition) methods. To reduce absorption, diamond anvils can be partially drilled [77].

The study of the electrical transport characteristics (Fig. 4 a,f) allows one to determine the type of hydride conduction, critical parameters of the superconducting state such as critical temperature ($T_C$), critical electric current density ($J_C$), upper critical magnetic field ($B_{C2}$), and electrical resistance at ambient conditions. In some cases, the Debye temperature can also be estimated using the Bloch–Grüneisen formula [78,79]. The compressibility $V^{-1}dV/dP$, calculated from the experimental equation of state, can be used to obtain some mechanical parameters and using theoretical models, to estimate the Debye temperature.

Polyhydrides can be considered as intermetallides formed by metals and metallic hydrogen. One of the effective approaches to search for their structures is compare them with those of known binary intermetallides formed by atoms with significantly different radii. Such an approach was successfully applied in the theoretical study of clathrate $Li_2MgH_{16}$ (predicted $T_C$ up to 473 K [32]) and in the experimental discovery of $Eu_8H_{46}$ [43] and $Ba_8H_{46}$ [80], which have a large number of prototypes such as $Ba_4Si_{23}$, $Ba_4Ge_{23}$, $Cs_4Sn_{23}$, etc.

Of interest is an entirely mathematical approach to the study of the structures of inorganic compounds under pressure proposed by R. Koshoji and colleagues [81,82] who studied the closest packing of spheres of different radius in three-dimensional Euclidean space. It is known that at high pressures, the packing density of atoms is one of the decisive factors in the stabilization of chemical compounds. For packings of two types of spheres ($A$ and $B$) the authors [81,82] found that with a large radius ratio $r_A/r_B$, the space is most optimally filled with clathrate structures at ratios $A:B$ = 1:12, 1:10, 1:9, and 1:6. Polyhydrides of many elements do have such stoichiometries and crystal structures. Unfortunately, the exact ratio of the effective radii of the atoms of hydrogen and the hydride forming element depends both on pressure and the atomic charges, which prevents the results of these works from being applied directly. Nevertheless, they provide a mathematical basis for understanding the formation of polyhydrides at high pressures.

Polyhydrides of many elements do have such stoichiometries and crystal structures. Unfortunately, the exact ratio of the effective radii of the atoms of hydrogen and the hydride forming element depends both on pressure and on the atomic charges, which prevents the results of these works from being applied directly. Nevertheless, they provide a mathematical basis for understanding the formation of polyhydrides at high pressures.

# 4. Experimental techniques

In 1959, diamond anvils were first used to create ultrahigh pressures in special cells [83]. The use of diamond, which is optically transparent (up to 220 nm, bandgap 5.5 eV) and the hardest of known materials, has opened up wide opportunities both for increasing the range of investigated pressures and for applying optical and X-ray diffraction methods for studying materials. In 1978, a significant improvement was made to the diamond anvil by adding a series of bevels in the vicinity of the culet to smoothen its shape [84].

This improvement allowed one to systematically reach pressures of 100–200 GPa and perform routine experiments with many materials. In particular, lots of works in 1970–2000 explored the behavior of pure elements under pressure. In terms of superconductivity, 22 elements were found to adopt the superconducting state under pressure, in addition to the previously known 31 elements superconducting at ambient pressure [85]. These discoveries led to the understanding that an increase in pressure promotes the emergence and enhancement of superconducting properties. The critical temperatures often increase at compression, and the behavior of $T_C(P)$ function is nonlinear and often "presents surprises" (e.g., in NbTi [86]).

**The design of the diamond anvil cells includes**:

(1) A gasket which is located between the diamond anvils. The gaskets are ceramic or metal plates, which can be made of BN, MgO, CaF2, Re, W, Al, Be, etc.. The gasket has a hole, which is a working volume where the sample under study is located. It also serves as the walls of the cell, where high pressure is created.

(2) Pressure transmitting medium to the sample (H$_2$, Ar, Ne, He, organic liquids, ammonia borane NH$_3$BH$_3$, etc.).

(3) Pressure sensor (ruby luminescence, X-ray diffraction from gold or platinum). The pressure can also be estimated from the edge of the Raman signal from the diamond.

(4) An insulating layer (Al$_2$O$_3$, 5–100 nm) applied to the anvils to protect them from aggressive media (hydrogen, helium, fluorine, etc.) and to thermally insulate the sample.

(5) An electrode system, usually multilayered, which is used to supply and read the electrical signal from the sample (Au, Mo, Au/Ta, B-alloyed diamond, etc.). Electrodes are formed using lithography, focused ion beam, magnetron sputtering or deposition from the gas phase (Physical Vapor Deposition, PVD).

(6) Diamond anvils (usually synthetic, Figure 4d), in which the shape and size of the culet mainly determine the maximum pressure attainable in the diamond anvil cell. Special shapes of diamond anvils allow generating pressures of up to a thousand gigapascals in an area of several microns [87]. To improve the performance, the culet surface of diamond anvils can be modified using focused ion beam etching (Xe Focused Iron Beam, FIB). Because diamond anvils cannot be unloaded without partial cracking due to jamming effects when pressures reach 70–80 GPa, experiments with pressures above 1 megabar (100 GPa) almost always require their replacement or regrinding and repolishing.

(7) Bases for diamond anvils (seats), which transmit and distribute the force from the cell to the anvils with minimal deformation. As a rule, seats are usually made of tungsten carbide and boron nitride. The bases with a conical anvil seat have the best characteristics [ 88].

(8) The diamond cell cylinder and piston, made of beryllium or titanium bronzes, or of special nonmagnetic steel, and screws and springs to create and smoothly transmit the force (Fig. 4 c,f). The cell material should be as hard as possible, nonmagnetic, and having minimum thermal expansion. Suitable alloys are beryllium and titanium bronzes and NiCrAl alloys, which are very difficult for milling and lathing.

The most important step in obtaining superhydrides is the synthesis under laser heating. As a source of hydrogen, the authors of this article systematically use the solid $NH_3BH_3$ (ammonia borane, or AB) [88-91], which decomposes into hydrogen and amorphous polymer $[NBH_x]_n$ at temperatures above 200–250 ºC [92,93]. In principle, hydride synthesis can be carried out at a sufficiently low temperature of 250–400 ºC, although heating to 1000–1500 ºC is more commonly used, at which the sample heats and saturates with hydrogen more uniformly. The heating of the metal target accelerates its reaction with hydrogen to form a hydride stable at this temperature and pressure. It is important to fix the sample between the anvils so that it does not touch them. For this, a sandwich structure is created, AB/sample/AB or AB/sample/electrodes). The fact is that diamond has an extremely high thermal conductivity, and if the sample is pressed against one of the anvils, its effective laser heating becomes impossible.

## 5. Peculiarities of the superconducting properties of polyhydrides

High-temperature superconductivity in various hydrides under pressure, predicted by Neil Ashcroft [94], was then theoretically studied by DFT methods and experimentally discovered in many compounds. To date, more than 90–95% of the works on hydrides are still theoretical studies. It is important that in almost all cases, *ab initio* calculations resulted in an overestimation of the critical temperature of superconductivity (Table 1). The reason was the disregard for the anharmonicity of the vibrations of the hydrogen sublattice as well as a possibly higher effective Coulomb pseudopotential µ* to 0.2 (usually, µ* = 0.1–0.15 is assumed in calculations).

Table 1. Highest critical temperatures obtained experimentally and theoretically in the harmonic approximation (at µ* = 0.1) of some hydride superconductors.

| Compound | Experimental pressure, GPa | Estimated $T_C$, K | Experimental $T_C$, K |
|---|---|---|---|
| $Im\bar{3}m$-$H_3S$ | 150 | 200 [15] | 203 [5] |
| $Fm\bar{3}m$-$LaH_{10}$ | 160 | 286 [18,19] | 250 [7] |
| $P6_3/mmc$-$YH_9$ | 200 | 253 [19] | 243 [31] |
| $Im\bar{3}m$-$YH_6$ | 170 | 270 [96] | 224 [30] |
| $Fm\bar{3}m$-$ThH_{10}$ | 170 | 160–193 [27,97] | 161 [27] |
| $P6_3/mmc$-$UH_7$ | 70 | 46 [26] | 8 [45] |
| $F\bar{4}3m$-$PrH_9$ | 150 | 56 [36] | 6 [25] |
| $P6_3/mmc$-$CeH_9$ | 110 | 117 [28,29] | ~90 [98] |
| $Fm\bar{3}m$-$CeH_{10}$ | 100 | 168 [99] | ~115 [98] |
| $c$-$SnH_x$ | 190 | 81–97 [100] | 76 [101] |
| $PH_x$ | 200 | ~100 [102] | 100 [103] |
| $Pm\bar{3}n$-$AlH_3$ | 110 | >24 [104,105] | <4 [105,106] |
| $Im\bar{3}m$-$CaH_6$ | 170 | 220–235 [107] | 215 [108] |

\* The presented theoretical $T_C$ values were obtained prior to the publication of experimental works. The comparisons shown in the table are illustrative, as it is difficult to find data for the same pressure.

The superconducting properties of metallic and covalent hydrides differ in a number of aspects. One of the features of covalent hydrides is nonlinear temperature dependence of the electrical resistance in the normal state, observed for both $H_3S$ [5] and $CSH_x$ [37]. This prevents even an approximate estimation of the Debye temperature using the electrical resistivity in the ambient state and the Bloch–Grüneisen formula [78,109]. Another feature of covalent hydrides is a relatively low upper critical magnetic field $H_{C2}$. Thus, for the room-temperature superconductor CSH$_x$ with $T_C$ = +(13–15ºC), the extrapolated value of $H_{C2}(0) \sim 70$ T [37], whereas for the "weaker" superconductors LaH$_{10}$, YH$_6$, and YH$_9$ with $T_C < 250$ K it exceeds 100–120 T. For comparison, large values of $H_{C2}(0)$ — up to 300 T on extrapolation — can be achieved only in some iron-containing pnictides, for example in NdFeAsO$_{0.82}$F$_{0.18}$ ($T_C$ = 49 K) [110].

The results of Errea and Bergara [106, 111-113] led to the understanding of a large contribution of anharmonicity of vibrations of the hydrogen sublattice to the thermodynamic stability and superconductivity of polyhydrides and, to a lesser extent, polydeuterides. Using the Stochastic Self-consistent Harmonic Approximation (SSCHA) method, the authors [106, 111-113] could answer many questions. Among them are the question why $T_C$ of palladium deuteride PdD is higher than that of the corresponding hydride PdH, as well as the question about the unexpected stability of decahydride LaH$_{10}$, which should decompose below 210 GPa in the harmonic approximation but exists at pressure decrease down to 140–145 GPa in experiment. The analysis of the anharmonic corrections shows that in many cases the critical temperature in hydrides lowers by 20–25 K, the electron–phonon interaction coefficient decreases by 20–25% (due to anharmonic hardening of soft phonon modes), whereas the logarithmic frequency increases by 40–50% (300–350 K) in comparison with the harmonic approximation (Table 2). This decrease in $T_C$ is critical for compounds with low predicted $T_C$ (e.g., AlH$_3$ [105]), in which anharmonic effects practically suppress superconductivity.

Table 2. Comparison of superconducting state parameters of hydrides in harmonic (h) and anharmonic (ah) approximations.

| Componnd (pressure, GPa) | $\lambda^h$ | $\lambda^{ah}$ | $\omega_{\log}^h$, K | $\omega_{\log}^{ah}$, K | $T_C^h$, K | $T_C^{ah}$, K | $T_C^{\exp}$, K |
|---|---|---|---|---|---|---|---|
| $Im\bar{3}m$-H$_3$S (200) [111] | 2.64 | 1.84 | 1049 | 1078 | 250 | 194 | 190 |
| $Fm\bar{3}m$-LaH$_{10}$ (214) [112] | 3.42 | 2.06 | 851 | 1340 | 249 | 238 | 245 |
| $Im\bar{3}m$-YH$_6$ (165) [30] | 2.24 | 1.71 | 929 | 1333 | 272 | 247 | 224 |
| $Pm\bar{3}n$-AlH$_3$ (110) [106] | 0.95 | 0.52 | 485 | 1050 | 31 | 15 | <4 |
| PdH (0) [113] | 1.55 | 0.4 | 205 | 405 | 47 | 5 | 9 |
| The table illustrates the importance of taking into account anharmonicity when studying hydrides. | | | | | | | |

The main disadvantage of the SSCHA method is the computational complexity. The calculations of the anharmonic Eliashberg function for a single compound can take up to several months. A number of recent works have implemented another approach to account for anharmonic corrections. This method is based on the use of machine-learning potentials and molecular dynamics of polyhydride supercells containing ~1000 atoms [43,114-118]. This approach enables the calculation of the anharmonic spectral densities of phonon states at any given temperature in a few days and therefore to correct the phase diagram of compounds, the phonon spectrum, and the high-frequency part of the Eliashberg spectral function $\alpha^2 F(\omega)$.

The Migdal–Eliashberg theory [119,120] has one uncertain parameter responsible for the effective Coulomb interaction — the so-called Coulomb pseudopotential μ*, whose values are usually 0.1–0.15. Since even the anharmonic effects are often insufficient to explain the overestimation of $T_C$ in theoretical calculations, in several works [30.121,122], it was suggested that in hydrides under pressure, μ* can take much higher values of 0.2–0.5. The exact value of μ* significantly affects the critical temperature and other parameters of the superconductor, therefore getting this parameter correctly is important. Currently, the most common method for taking into account the effect of the Coulomb interaction on superconductivity is the so-called DFT method for superconducting compounds (Superconductor Density Functional Theory, SCDFT), which is based on solving the Kohn–Sham equations for the order parameter [123,124]. Successfully applied to many superconductors (Nb [125], MgB$_2$ [126], V$_3$Si [127], H$_3$S [128]), this method nevertheless gives underestimated $T_C$ (and thereby overestimated µ*) values for many superhydrides, for example, YH$_6$ [30], YH$_9$ [45], and LaH$_{16}$ [121]. There is some progress due to the recent introduction of a new exchange–correlation functional SPG 2020 [127], which better approximates the experimental values (Table 3). However, only a systematic application of the entirely anisotropic SCDFT method and comparison of predictions with experimental values of $T_C$ can help in the future to understand which values the Coulomb pseudopotential can really take in hydrides at high pressures.

Table 3. Comparison of the results of the critical temperature calculations within the DFT theory for superconductivity (SCDFT) and experiment.

| Compound (pressure in GPa) | $T_C$, K (LM 2005)* | $T_C$, K (SPG 2020) | $T_C^{\exp}$, K |
|---|---|---|---|
| $Im\bar{3}m$-H$_3$S (200) [128] | 180 | - | 190 |
| $Im\bar{3}m$-YH$_6$ (165) [30] | 156 | 181 | 224 |
| $P6_3/mmc$-YH$_9$ (200) [31,45] | 179 | 246 | 243 |
| $Fm\bar{3}m$-LaH$_{10}$ (214) [112] | 210 | - | 245 |
| $P6/mmm$-LaH$_{16}$ (200) [121] | 156 | - | 241** |
| *LM 2005 – exchange-correlation functional developed in [123] ||||
| ** Harmonic approximation, standard calculations using the Eliashberg equations. ||||

Another important correction to the superconducting transition temperature is the need to take into account the anisotropy of the superconducting gap. In most of the early works of 2010–2018, the Migdal–Eliashberg equations were solved in the isotropic approximation without accounting for anharmonicity and with empirical values of µ* = 0.1–0.15. However, in 2015 it was found that accounting for the anisotropy of the Fermi surface, electron–phonon and electron–electron interactions in the energy space in many hydrides leads to a ~20–30 K increase in the superconducting transition temperature [96,128-130] compared to isotropic calculations. It was shown that a hypothetical compound $Fm\bar{3}m$-YH$_{10}$ exhibits a significant gap anisotropy of Δ ± 5 meV, whereas $Im\bar{3}m$-YH$_6$ has two superconducting gaps of 32 and 50 meV [94]. In the more recent work, Wang et al. [130] found that $Fm\bar{3}m$-LaH$_{10}$ also has a significantly anisotropic main superconducting gap, 46 ± 5 meV, and a small additional gap $\Delta_2 \approx 6.2$ meV. The anisotropy of the electron-phonon interaction should now be taken into account in all cases, regardless of the superhydride structure. For many hydrides, solution of the anisotropic Migdal–Eliashberg equations adds approximately 20–30 K to $T_C$ found within the isotropic theory.

An important feature of superconductivity is the existence of a critical electric current density $J_C$. As we first show in several our works [27,30,118], in hydrides the extrapolated critical current density $J_C(0)$ reaches very large values, from 10 to 100 kA/mm$^2$, which are comparable or exceed those of all currently known types of superconductors. When estimating the critical current density, attention must be paid to the thickness of the sample placed between the diamond anvils.

The thickness of the sample should not exceed the distance between the diamond culets, determined by interference of visible light, and at pressures above 100 GPa is about 1 µm. In this case, a sample diameter is about 20–40 µm, and a current at a liquid helium temperature is several amperes. The possibly labyrinthine current flow can be additionally considered based on theoretical calculations of the normal state resistance of hydrides using the EPW package [131-133]. The calculations show that due to the strong electron–phonon interactions, the electrical resistivity of hydrides in their normal state is very high, being at the level of such materials as mercury, constantan, and nichrome. Using the van der Pauw formula [134,135] for estimation, the effective thickness of the YH$_6$ or LaYH$_{20}$ hydride samples of 0.5–0.75 µm can be obtained, which further increases the critical current density estimate in superhydrides.

In recent years, studies of hydrides is increasingly focused on ternary systems such as C–S–H [53, 136] and Y–Pd–H [74], in this regard, a few words should be said about the problem of superconductivity in such systems. The authors [118] recently showed that during the synthesis of the La–Y–H system from the La–Y alloy both metal atoms are randomly distributed among common positions in the metal sublattice.

In the X-ray diffraction pattern, a set of lines characteristic of pure binary hydride of one of the components (e.g., $LaH_{10}$) with an altered volume of the sublattice is observed and no peak splittings or superstructure reflections are detected. Because of this disorder, the width of the superconducting transition of compounds increases significantly — up to several tens of degrees (10–50 K). This broadening is an expected effect for all complex disordered systems, which will narrow the field of potential applications of multicomponent hydride superconductors in the future.

Table 4 shows the reproducibility of the temperature measurements of resistive (and sometimes magnetic) superconducting transitions in different hydrides. These experiments were performed using various initial materials containing different impurities. Cells were loaded in an inert atmosphere and in the air, and the hydrogen source was both ammonia borane and hydrogen gas. Various authors used different equipment for laser heating (and even synthesis by keeping samples in the hydrogen atmosphere for a long time), cryostats and thermometers, etc. Resistive transitions in hydrides are reproduced with a good accuracy of 10–15 K (~5%). At the same time, these results make unlikely the idea that insignificant impurities of carbon, boron, and nitrogen in narrow concentration limits are able to increase dramatically the critical temperature of superconductivity in hydrides [56,57]. The insignificant effect of nonmagnetic impurities on superconductivity can be investigated directly; such a study was performed, for example, in the group of W. Chen, X. Huang, and T. Cui [137] for carbon-doped $LaH_{10}$/C or aluminum-doped $LaH_{10}$/Al. They showed that in this case $T_C$ decreases by only ~5–10 K.

Table 4. Reproducibility of the measurements of the superconducting transition temperature by the drop in the electrical resistance (in some cases, by the jump in the magnetic susceptibility) obtained by different scientific groups.

| Compound | Maximum experimental $T_C$, K | Scientific group |
|---|---|---|
| $Im\bar{3}m$-$H_3S$ | 204<br>190<br>183 | Mainz (A. Drozdov, ... M. Eremets) [5]<br>Osaka (M. Einaga, ... K. Shimizu) [138]<br>Jilin University (X. Huang et al. ) [139] |
| $Fm\bar{3}m$-$LaH_{10}$ | 260<br>250<br>250<br>245 | Illinois (M. Somayazulu, ... R. J. Hemley) [8]<br>Mainz (A. Drozdov, ... M. Eremets) [7]<br>Beijing (Fang Hong et al., Institute of Physics, CAS*) [140]<br>Jilin University (W. Chen, ... X. Huang) |
| $Im\bar{3}m$-$YH_6$ | 224<br>227<br>211 | Moscow (I. Troyan et al. [30])<br>Mainz (P. Kong, ... M. Eremets [31])<br>Jilin University (W. Chen, ... X. Huang) |
| $c$-$SnH_4$ | 75<br>71 | Beijing (Fang Hong et al., Institute of Physics, CAS*) [101]<br>Moscow (I. Troyan) |
| $P6_3/mmc$-$YH_9$ | 243<br>237<br>230 | Mainz (P. Kong, ... M. Eremets [31])<br>Moscow (I. Troyan, D. Semenok et al. )<br>Bristol (J. Buhot et al. [141]) |
| $Im\bar{3}m$-$CaH_6$ | 215<br>195–210 | Jilin University (L. Ma, ... Y. M. Ma) [108]<br>Beijing (Z. W. Li et al., Institute of Physics, CAS*) [142] |
| * Institute of Physics of the Chinese Academy of Sciences. | | |

## 6. Criticism of hydride superconductivity

The discovery of a large number of superconducting hydrides ($H_3S$, $LaH_{10}$, $ThH_{10}$, $YH_6$, $YH_9$, and $CSH_x$) has attracted the attention of researchers in various fields. However, J. E. Hirsch, F. Marsiglio, M. Dogan, and M. L. Cohen [72, 136-139] have expressed doubts about the existence of superconductivity in hydrides, as well as the possibility of describing it in terms of the electron–phonon coupling mechanism. The authors of these works point to the small width of the superconducting transitions in hydrides, the insufficient broadening of SC transitions in an applied magnetic field, as well as the lack of clear evidence of diamagnetic shielding and the Meissner–Ochsenfeld effect [147] for hydrides. The latter is not surprising, since the available instrumental techniques for studying microscopic samples at ultrahigh pressures are limited to spectroscopic and X-ray diffraction methods. Along with this, electrode sputtering techniques and electrical measurements in diamond anvil cells are relatively well developed.

The sensitivity of detecting a superconducting transition by the van der Pauw four-contact method [134,135] is proportional to $L/S$ [m$^{-1}$], where $L$ is the characteristic size (diameter) of the sample, $S$ is the average cross-sectional area, whereas the magnetic field change in the vicinity of the sample is proportional to its volume $L \times S$ [m$^3$]. Therefore, resistive measurements are well suited for micron-sized samples; whereas magnetic measurements with such samples are technically very difficult [148]. Recently, promising new methods for studying diamagnetic shielding have been developed. These are based (i) on detection of fluorescence of nitrogen NV-centers created on the surface of a diamond anvil (Fig. 5c), and (ii) on the use of high-frequency current passing through single-turn coils sputtered on the culet (working surface) of a diamond anvil in the immediate vicinity of the sample. (Fig. 5 a,b).

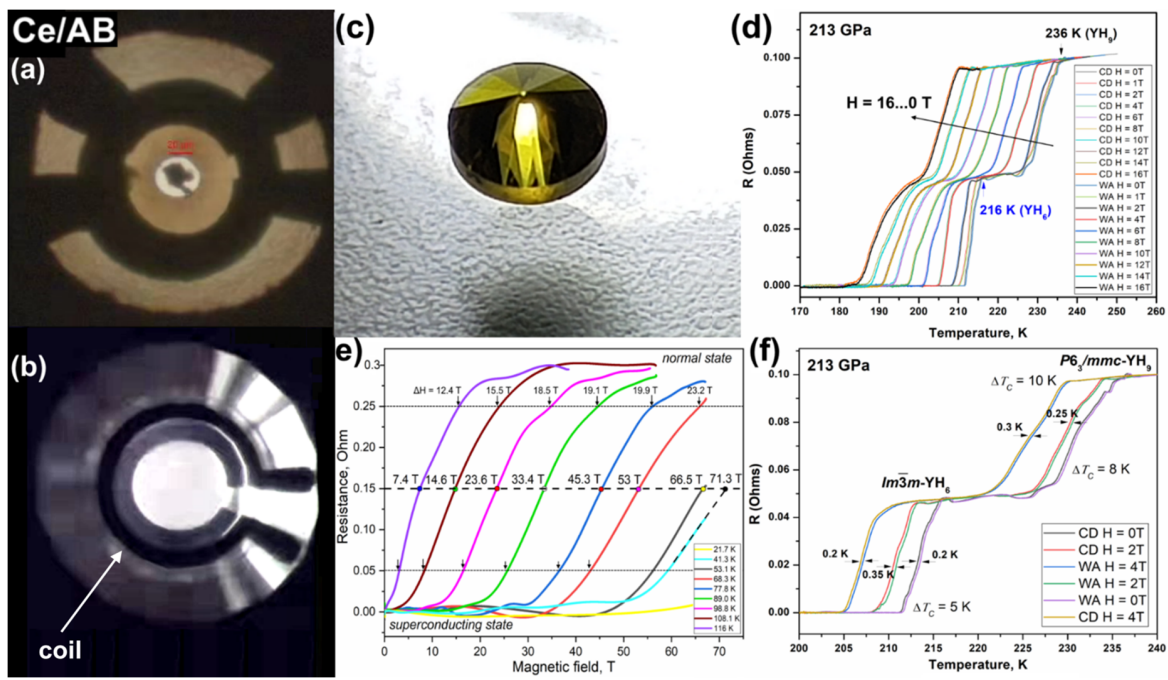

**Figure 5** (color online). (a, b) Single-turn coils sputtered on diamond anvils in the immediate vicinity of the sample after loading and before loading, respectively. (c) The nitrogen-doped diamond anvil with NV centers. (d) Broadening of superconducting transitions in a two-phase $YH_6 + YH_9$ sample in a magnetic field at pressure 213 GPa. (e) Broadening of superconducting transitions in a $(La,Nd)H_{10}$ sample containing 7–9 mol % of neodymium in strong pulsed magnetic fields up to 70 T [45]. (e) Hysteresis of superconducting transitions in a two-phase $YH_6 + YH_9$ sample at 213 GPa during cooling (CD) and heating (WA).

The critical attitude to the small width of SC transitions [54] is primarily related to the measurement of the $CSH_x$ hydride by Snider et al. [31]. The results of these studies revealed a shift of the superconducting transition by 20 K practically without broadening in a magnetic field of 9 T. We note that the doubts of J. Hirsch and F. Marsiglio about the width of superconducting transitions may be partially resolved. It should be taken into account that metal superhydrides have very high values of the upper critical magnetic field ($B_{C2}(0)$ exceeds 100–150 T), whereas most studies in a permanent magnetic field are related to weak fields ($B_{C2}(T)/B_{C2}(0) \sim 0.1$), when the broadening of SC transitions is insignificant.

In addition, the samples often contain impurities of lower hydrides, which give multistep transitions to the superconducting state. Finally, the samples under study are not single crystals, but most likely consist of grains. Therefore, their magnetic characteristics may be more similar to those of granular superconductors. A more detailed analysis of the data for $YH_6$, $YH_9$, and $(La, Nd)H_{10}$ shows (Fig. 5 d – f) that the superconducting transitions in the hydrides we studied are significantly broadened in magnetic fields. Their initial width is determined by the quality of the sample. The width of the SC transitions in hydrides can be significantly reduced by repeating of laser heating and cooling cycles ("annealing"). This is especially true for ternary compounds, where this broadening is due to the random distribution of different metal atoms in the metal sublattice. It is worth of noting that the small broadening of the SC transitions cannot be considered an exclusive feature of superconducting hydrides. Many iron-containing superconductors, in particular those of class 11, show an extremely insignificant broadening even down to the lowest temperatures [149-151].

One of the most important arguments in favor of the electron–phonon nature of superconductivity in hydrides is the isotope effect, which is manifested in a decrease in the superconducting transition temperature when hydrogen is replaced by deuterium. This effect was observed in $H_3S$ [5], $LaH_{10}$ [7], $YH_6$ [30], $YH_9$ [31], $CeH_{9-10}$ [98], $(Pd,Y)H_x$ [74], and in a number of other compounds. In all cases, the isotope coefficient $\alpha = -\ln(T_C)/\ln(M)$, where $M$ is the mass of the atom, was in the range of 0.3–0.6, in reasonable agreement with the theoretical prediction. A certain difficulty of the analysis is due to the fact that the chemistry of deuterides does not completely coincide with that of hydrides; the stability limits and the distortion regions of hydride and deuteride structures with respect to pressure differ even more. For this reason, the comparison of $T_C$ for hydrides and deuterides at the same pressure is sometimes incorrect, since their crystal structures may be different. Another factor complicating the comparison is a significantly smaller effect of anharmonicity on superconductivity in deuterides.

In general, deuterides exhibit the same properties as hydrides. The superconducting transition is shifted depending on the applied magnetic field. The upper critical field $B_{C2}(0)$ in deuterides, as a rule, is significantly lower than that in hydrides. There is a critical current, whose value also depends on the magnetic field. With a decrease in pressure, first the crystal structure of superdeuterides distorts, the critical temperature of the SC transition decreases markedly, then the compound decomposes with the formation of lower deuterides and $D_2$.

Experiments on observing diamagnetic screening in hydrides and their criticism have been of particular interest recently [61,139,143,152-155]. The unexpectedly high ability of hydrides to screen an external magnetic field is actively discussed. In particular, in one of the first works of Hirsch [143], the existence of superconductivity in $H_3S$ was questioned. However, as shown by D. M. Gokhfeld, it is necessary to correctly take into account the penetration of the magnetic flux into the sample. In a type II superconductor ($H_3S$), Abrikosov vortices and a magnetic field in the center of the sample are absent until the external field is less than the total penetration field $H_p$. In the critical state model [156]

$$H_p \sim J_c *a,$$

where $J_c$ is the critical current and $2a$ is the sample size in the direction perpendicular to the external field.

The field $H_p$ for this sample can be greater than the external field $H_{ext}$ = 0.68 T (determined from measurements [152]), as well as the field $H_{edge}$ ~ 4.3 T (expected at the edges of the sample plate due to the demagnetizing factor $H_{edg} = H_{ext}/(1-N)$). Assuming that the penetration depth of the magnetic flux in the experiment [152] was less than 5 μm, the estimate of $J_c$ from the depth of the screened region $J_c = H_{edge}/5$ μm gives a reasonable value of $J_c$ ~ 6.8 × $10^7$ A cm$^{-2}$, comparable to the critical density of intragranular currents in cuprate HTSCs [157].

Meanwhile, the lower critical field $H_{c1}$ in H$_3$S is apparently much less than 0.68 T. In this field range ($H_{c1} < H < H_p$), the distribution of Abrikosov vortices in the sample should be inhomogeneous, as in all type II superconductors, and the magnetic flux density decreases from the edges of the sample to the center. Therefore, the formulas for a uniform field derived in [143], are not applicable to this experiment. When evaluating the critical current density, one should also take into account the current circulation in a layer equal to a sample thickness of 5 μm (along the field direction) and a depth equal to the vortex penetration depth [158] (in the plane perpendicular to the field), rather than the penetration depth of the magnetic field λ. Using a layer cross section of 5 × 5 μm$^2$, we obtain $J_c$ = 6.4 × $10^7$ A cm$^{-2}$, which agrees with the results of recent measurements [30, 118], and is more than an order of magnitude lower than the estimate in [143]. Thus, for the screening effect established in Ref. [152], the values of $H_{c1}$ and $J_c$ for H$_3$S are quite consistent with similar parameters for other superconductors. The Hirsch et al. arguments [143] about "nonstandard superconductivity" in H$_3$S, in our opinion, are based on an overestimated value of $H_{c1}$ and therefore are incorrect.

When analyzing the recent work of Minkov et al. [154], where the expulsion of the magnetic field from LaH$_{10}$ and H$_3$S samples was studied using a SQUID magnetometer, one should take into account that the hydride samples are probably porous and consist of microscopic grains (~ 0.05-0.5 μm). In this case, the demagnetization factor should be calculated for a random packing of spherical particles and ranges from 0.33 to 0.5 [159,160]. The magnetic field penetrates the sample between the individual grains. Therefore, no change in the magnetization of the sample is observed at temperatures around $T_C$ upon cooling in the magnetic field (Field Cooling, FC). Thus, the penetration fields found by the authors of the works [159,160] $H_P(0)$) = 96 mT for H$_3$S and 41 mT for LaH$_{10}$ are the lower bound of $H_{c1}(0)$. Whereas a more realistic estimate is $H_{c1}(0)$ ~ $H_P(0)/(1-N)$ = (1.5-2) $H_P(0)$.

***Thus, the following properties are observed in superconducting hydrides***:

1) The isotope effect upon replacement of hydrogen by deuterium, with α = 0.3–0.6.

2) The sharp drop in the electrical resistance (by a factor of $10^3$–$10^5$) to several micro-Ohms at a certain temperature ($T_C$) within a few Kelvins, which is the same in heating and cooling cycles.

3) The strong dependence of the critical temperature $T_C$ on the applied magnetic field, which is linear at low fields $H_{C2}$ ~ $|T-T_C|$. For many hydrides, this linear $H_{C2}$ (T) dependence persists down to the lowest temperatures.

4) The presence of a critical current $I_C$, which depends on the applied magnetic field and temperature.

5) The dome-shaped dependence of the critical temperature on pressure. At low pressures, it corresponds to distortion of a high-symmetry crystal structure. At high pressures, it corresponds to a decrease in the electron-phonon coupling constant due to the anharmonic hardening of soft phonon modes.

6) The temperature broadening of the superconducting transitions in a magnetic field (especially noticeable in strong pulsed fields).

7) The temperature broadening of the superconducting transitions in ternary hydrides (transition width up to 30–50 K) due to disordered structure.

8) The significant suppression of $T_C$ in hydrides by paramagnetic impurities (e.g., 1 atom % of Nd leads to $\Delta T_C \approx -10$ K) and an insignificant effect of nonmagnetic impurities (C, Al, Be) on $T_C$.

9) The diamagnetic screening, probably registered for $H_3S$, $CSH_x$, $LaH_{10}$, and $CeH_9$ in several experiments (see also the recent work Ref. [154]).

10) The probable approaching of the reflectivity to unity in the infrared range at incident radiation energies less than $\sim 2\Delta = 73$ meV. Here the criticism [161] of such experiments must be taken into account.

All these properties find the most consistent explanation in terms of superconductivity, which is also expected from *ab initio* calculations. So far, only single deviations are known in the behavior of hydrides from the Bardin–Cooper–Schrieffer–Migdal–Eliashberg theory (e.g., the linear dependence of $H_{C2}(T)$ over the entire temperature range or anomalously low $T_C$ for $YH_6$). At the same time, no alternative interpretations explaining the entire set of the observed phenomena have been proposed by the authors of critical articles.

## 7. Future research directions

However, the rationale behind the critique of J. Hirsch and F. Marsiglio (see Section 6) is that hydrides should be investigated in more detail. At the moment, the basic parameters of the electron–phonon interaction (EPI constant $\lambda$) in these compounds, superconducting gap $\Delta$, Eliashberg function $\alpha^2 F(\omega)$, logarithmic frequency $\omega_{\log}$, and Coulomb pseudopotential $\mu^*$ are known mainly from the first-principles (*ab initio*) calculations. Obviously, future studies will have to fill this void. There are several promising approaches for experimental investigation of the superconducting state parameters of polyhydrides that can be realized in high-pressure diamond anvil cells (DAC).

I. Femtosecond reflection spectroscopy. This method can allow direct experimental determination of EPI constant from the relaxation rate of the electron temperature, as was done for simple metals and intermetallic compounds [162]. Optical non-linearity of diamond, which leads to defocusing of the femtosecond pulse, poses a problem here.

II. Infrared reflection spectroscopy in a wide energy range at low temperatures. This method allows direct determination of the superconducting gap $\Delta$ and its temperature dependence $\Delta(T)$. This was already used in the study of $H_3S$ in 2017 [76]. The disadvantage of the method is the need for large samples, 70–150 μm, which leads to a limit on the maximum pressures (< 175 GPa) in the cells. Nevertheless, $LaH_{10}$, $YH_6$, $ThH_{10}$, and $CeH_{9-10}$ are the primary targets for this method.

III. Pulsed ultrahigh magnetic fields (up to 70–80 T, and much higher using explosive magnetic flux compression). This technique began to be applied to diamond anvil cells relatively recently (see works of 2019–2020 on $H_3S$ and $LaH_{10}$ [163,164]). The method allows to reduce significantly the uncertainty in extrapolating the upper critical magnetic field $H_{C2}(0)$, plot a magnetic phase diagram down to the lowest temperatures, and verify which model works best for the $H_{C2}(T)$ dependence. Difficulty in this case is the necessity to perform measurements at high frequencies (3–100 kHz) in miniature DAC ($d$ = 15 mm) made of special steel. The method imposes serious requirements to minimize parasitic capacitance and inductance in the electrode system of the DAC.

The currently available pulsed magnetic fields (70–80 T) are still not strong enough to completely suppress superconductivity in the most interesting superhydrides, for which $H_{C2}(0)$ exceeds 120–140 T. Therefore, this technique is most effective for compounds with low $T_C \sim 100–150$ K, whereas for H$_3$S and LaH$_{10}$, the obtained $H_{C2}(T)$ dependences remain linear even in the strongest available magnetic fields.

IV. Andreev reflection spectroscopy and microcontact spectroscopy [165] are promising research methods that allow determination of not only the value of the SC gap $\Delta(T)$ but also of its anisotropy when there are several gaps simultaneously. This method was recently successfully used to establish an anisotropic nature of the superconducting gap in metallic yttrium at high pressure [166]. The authors found that at high pressures yttrium has two superconducting gaps, around 3.6 and 0.5 meV, with $2\Delta/(k_B T_C)$ ratio reaching 8.2 (where $k_B$ is the Boltzmann constant), which is in favor of superconductivity with strong coupling. The difficulty of this kind of study is that the Andreev contact must be nano-sized, which is hard to monitor when the sample is compressed and heated in the diamond anvil cells.

Thus, the design of future experimental studies of hydride superconductivity should include single crystal X-ray diffraction in high pressure DAC; reflectance UV-Vis-IR spectroscopy in the ultraviolet (UV), visible (Vis) and infrared (IR) spectral ranges with the determination of the superconducting gap value $\Delta(T)$; resistive measurements in a wide frequency range up to 10–100 kHz in steady and strong pulsed magnetic fields with detection of $H_{C2}(T)$, $J_C(T)$, magnetoresistance, Hall effect, and Andreev reflection; as well as measurements of magnetic susceptibility, $H_{C1}(T)$, and magnetic ordering character by X-ray magnetic circular dichroism (XMCD) in a magnetic field in lanthanoid superhydrides (Nd, Sm, Gd, Eu, etc.).

From the theoretical point of view, future works will include thermodynamic calculations with machine-learning interatomic potentials at finite temperatures, taking into anharmonicity into account and allowing one to study large systems (100–150 atoms and more). In addition, calculations of both superconductivity and resistivity at ambient conditions and electron–phonon coupling anisotropy using *ab initio* calculations of the Coulomb interaction contribution using the SCDFT method will have to be carried out.

More complex experiments on the use of superconducting hydrides include the fabrication of conducting structures on the surface of a diamond anvil, for example, the fabrication of S – N – S interfaces with a insulation gap of about 1–10 nm between superhydride electrodes and SQUID magnetometers, the fabrication of multilayer interfaces using layer-by-layer deposition of various metals and oxides, placing microthermometers and microheaters on diamond to measure the jump in heat capacity, as well as the fabrication of micro-rings from superhydrides for studying magnetic flux trapping. It is also important to determine positions of hydrogen (deuterium) atoms at least in some of superhydrides, stable at low pressures (ThD$_4$, UD$_{5–8}$, CeD$_{8–10}$), using neutron diffraction to verify the results of theoretical calculations.

## 8. Conclusions

For six years of research (2015–2021) since the unique properties of H$_3$S were first predicted, and then experimentally discovered [5], polyhydrides proved to be a new class of superconducting materials with record critical parameters. Undoubtedly, many more exciting discoveries can still be made in this field. Hydrogen is an ideal element for high-temperature superconductivity with the electron–phonon mechanism. To implement this, it is only required to find polyhydrides, which would maintain record high critical temperatures at low pressures. Hybrid metastable materials combining both covalent bonds and a hydrogen atom sublattice stabilized by metals have a great potential in this field.

Progress in hydride superconductivity would not be so fast and bright without the well-developed Migdal–Eliashberg theory of strong electron–phonon interaction, methods of evolutionary search for thermodynamically stable crystal structures (USPEX), as well as extremely successful software packages Quantum Espresso and EPW for *ab initio* calculations of critical parameters of superconducting states of crystals. The developers of the SSCHA package should also be noted for their contribution to understanding the importance of anharmonic effects in hydrides at ultrahigh pressures.

## Acknowledgements

This work was supported by the Ministry of Science and Higher Education within the State assignment of Federal Science Research Centre "Crystallography and Photonics" RAS. I.A.T. thanks the Russian Science Foundation (RSF Project № 22-12-00163) and A.R.O. thanks the Russian Science Foundation (RSF project № 19-72-30043) for financial support of the projects. D.V.S. thanks the Russian Foundation for Basic Research (RFBR project № 20-32-30043). A.V.S. thanks the Russian Science Foundation (RSF project № 22-22-00570) for financial support of the study. The work of A.V.S., O.A.S. and V.M.P. was carried out within the framework of the state assignment of the P. N. Lebedev Physical Institute of the Russian Academy of Sciences (project No. 0023-2019-0005 "Physics of High-Temperature Superconductors and New Quantum Materials"). The authors are grateful to Dr. D. Gochfeld for helpful comments on the details of the observation of the Meissner effect in $H_3S$. The authors would like to thank Dr. Wuhao Chen (Jilin University, China) for providing photos of diamond anvils, to Dr. Niu Haiyang (Northwestern Polytechnical University, China) for help in calculating the C-S-H system, as well as to Igor Grishin (Skoltech) for proofreading and editing the Russian version of the article, and Marianna Lyubutina (FSRC "Crystallography and Photonics" RAS) for correcting the text of the English version of the article.